\newcommand{\stdpack}{
  \usepackage{amssymb}
  \usepackage{amsmath}
  \usepackage{eucal}
  \usepackage[final]{graphicx}
  \usepackage{psfrag}
  \usepackage{fancyhdr}
  \renewcommand{\headrulewidth}{0pt}\lhead{}\cfoot{}\rfoot{\thepage}

  \newcommand{\draft}{\usepackage[light,first]{draftcopy}\draftcopyName{draft}{350}}
  \newcommand{\labels}{\usepackage{showlabels}}
  \newcommand{\maple}{\usepackage{maple2e}}
  \newcommand{\makeidx}{\usepackage{makeidx}\makeindex}
}
\newcommand{\std}[1]{
  \stdpack
  \usepackage{a4,a4wide}
  \newcommand{\blockindent}{3ex}
  \renewcommand{\baselinestretch}{#1}
  \renewcommand{\arraystretch}{1.2}
  \hoffset -1cm  \addtolength{\textwidth}{2cm}
  \voffset -0cm  \addtolength{\textheight}{0cm}

  \usepackage{./mt}
  \columnsep 5ex
  \parindent 3ex
  \parskip 1ex
  \macros
  \pagestyle{fancy}
  \bibliographystyle{/home/mt/bibtex/chicago}
\renewenvironment{abstract}{\paragraph{Abstract}\begin{rblock}\small}{\end{rblock}}
}

\newcommand{\article}[2]{
  \documentclass[#1pt,twoside,fleqn]{article}\usepackage{chicago}\std{#2} }
\newcommand{\nips}{
  \documentclass{article} \usepackage{nips01e,times} \stdpack\macros }
\newcommand{\ijcnn}{
  \documentclass[10pt,twocolumn]{/home/mt/usr/tex/ijcnn}
  \stdpack\macros
  \bibliographystyle{abbrv} 
}
\newcommand{\foga}{
  \documentclass{article} 
  \stdpack
  \usepackage{/home/mt/usr/tex/foga}
  \macros
}

\newcommand{\book}[2]{
  \documentclass[#1pt,twoside,fleqn]{book}\usepackage{chicago}\std{#2} }

\newcommand{\foils}[1]{
  \documentclass[12pt,fleqn]{article}
  \std{#1}
  \voffset -1cm  \addtolength{\textheight}{2cm}
  \renewcommand{\footskip}{2cm}
  \begin{document}
  \large
}
\newcommand{\landfoils}[1]{
  \documentclass[fleqn]{article}
  \stdpack
  \renewcommand{\baselinestretch}{#1}
  \renewcommand{\arraystretch}{1}
  \setlength{\hoffset}{-3.5cm}
  \setlength{\voffset}{-3.5cm}
  \setlength{\textwidth}{27cm}
  \setlength{\textheight}{19cm}
  \parindent 0ex
  \parskip 0ex 
  \pagestyle{empty}
  \macros
  \begin{document}
  \huge
}
\newcommand{\landfolien}[1]{
  \documentclass[fleqn]{article}
  \usepackage{german}
  \stdpack
  \renewcommand{\baselinestretch}{#1}
  \renewcommand{\arraystretch}{1.5}
  \setlength{\hoffset}{-5cm}
  \setlength{\voffset}{-1.5cm}
  \setlength{\textwidth}{27cm}
  \setlength{\textheight}{19cm}
  \parindent 0ex
  \parskip 0ex 
  \pagestyle{plain}
  \begin{document}
  \huge
}

\newcommand{\addressCologne}{
  Institute for Theoretical Physics\\
  University of Cologne\\
  50923 K\"oln---Germany\\
  {\tt mt@thp.uni-koeln.de}\\
  {\tt www.thp.uni-koeln.de/\~{}mt/}
}

\newcommand{\homepage}{{\tt www.neuroinformatik.ruhr-uni-bochum.de/PEOPLE/mt/}}
\newcommand{\email}{{\rm mt@neuroinformatik.ruhr-uni-bochum.de}}

\newcommand{\address}{\small\it
  Institut f\"ur Neuroinformatik,
  Ruhr-Universit\"at Bochum, ND 04,
  44780 Bochum---Germany\\
  \email
}

\newcommand{\mytitle}{
  \thispagestyle{empty}
  \rhead{\it Marc Toussaint---\today}
  \hrule height2pt
  \begin{list}{}{\leftmargin2ex \rightmargin2ex \topsep2ex }\item[]
    {\Large\bf \thetitle}
  \end{list}
  \begin{list}{}{\leftmargin7ex \rightmargin7ex \topsep0ex }\item[]
    Marc Toussaint \quad\today

    \address
  \end{list}
  \vspace{2ex}
  \hrule height1pt
  \vspace{5ex}
}

\newcommand{\contents}{{\small \parskip 0ex \tableofcontents \parskip 2ex }}

\newcommand{\sepline}{
  \begin{center} \begin{picture}(200,0)
    \line(1,0){200}
  \end{picture}\end{center}
}

\newcommand{\partsection}[1]{
  \vspace{5ex}
  \centerline{\sc\LARGE #1}
  \addtocontents{toc}{\contentsline{section}{{\sc #1}}{}}
}

\newcommand{\intro}[1]{\textbf{#1}\index{#1}}

\newtheorem{definition}{Definition}
\newtheorem{statement}{Statement}
\newtheorem{theorem}{Theorem}
\newtheorem{hypothesis}{Hypothesis}
\newenvironment{remark}{\noindent\emph{Remark.}}{}
\newenvironment{example}[1][]{\begin{block}[Example {#1}]}{\end{block}~}

\newcounter{parac}
\newcommand{\para}{\refstepcounter{parac}{\bf [{\roman{parac}}]}~~}
\newcommand{\Pref}[1]{[\emph{\ref{#1}}\,]}

\newenvironment{block}[1][]{{\noindent\bf #1}
\begin{list}{}{\leftmargin\blockindent \topsep-\parskip}
\item[]
}{
\end{list}
}

\newenvironment{rblock}{
\begin{list}{}{\leftmargin\blockindent \rightmargin\blockindent \topsep-\parskip}
\item[]
}{
\end{list}
}

\newenvironment{keywords}{\paragraph{Keywords}\begin{rblock}\small}{\end{rblock}}

\newenvironment{colpage}{
\addtolength{\columnwidth}{-3ex}
\begin{minipage}{\columnwidth}
\vspace{.5ex}
}{
\vspace{.5ex}
\end{minipage}
}

\newenvironment{enum}{
\begin{list}{}{\leftmargin3ex \topsep0ex \itemsep0ex}
\item[\labelenumi]
}{
\end{list}
}

\newenvironment{cramp}{
\begin{quote} \begin{picture}(0,0)
        \put(-5,0){\line(1,0){20}}
        \put(-5,0){\line(0,-1){20}}
\end{picture}
}{
\begin{picture}(0,0)
        \put(-5,5){\line(1,0){20}}
        \put(-5,5){\line(0,1){20}}
\end{picture} \end{quote}
}

\newcommand{\inputReduce}[1]{
{\sc\hspace{\fill} REDUCE file: #1}
}

\newcommand{\inputReduceInput}[1]{
  {\sc\hspace{\fill} REDUCE input - file: #1}
  \input{#1.tex}
}

\newcommand{\inputReduceOutput}[1]{
  {\sc\hspace{\fill} REDUCE output - file: #1}
}

\newcommand{\todo}[1]{{\bf[#1]}}

\newcommand{\macros}{
  \newcommand{\0}{{\hat 0}}
  \newcommand{\1}{{\hat 1}}
  \newcommand{\2}{{\hat 2}}
  \newcommand{\3}{{\hat 3}}
  \newcommand{\5}{{\hat 5}}

  \renewcommand{\a}{\alpha}
  \renewcommand{\b}{\beta}
  \renewcommand{\c}{\gamma}
  \renewcommand{\d}{\delta}
    \newcommand{\D}{\Delta}
    \newcommand{\e}{\epsilon}
    \newcommand{\g}{\gamma}
    \newcommand{\G}{\Gamma}
  \renewcommand{\l}{\lambda}
  \renewcommand{\L}{\Lambda}
    \newcommand{\m}{\mu}
    \newcommand{\n}{\nu}
    \newcommand{\N}{\nabla}
  \renewcommand{\k}{\kappa}
  \renewcommand{\O}{\Omega}
    \newcommand{\p}{\phi}
    \newcommand{\ph}{\varphi}
  \renewcommand{\P}{\Phi}
  \renewcommand{\r}{\varrho}
    \newcommand{\s}{\sigma}
    \newcommand{\Si}{\Sigma}
  \renewcommand{\t}{\theta}
    \newcommand{\T}{\Theta}
  \renewcommand{\v}{\vartheta}
    \newcommand{\X}{\Xi}
    \newcommand{\Y}{\Upsilon}

  \renewcommand{\AA}{{\cal A}}
    \newcommand{\GG}{{\cal G}}
  \renewcommand{\SS}{{\cal S}}
    \newcommand{\TT}{{\cal T}}
    \newcommand{\EE}{{\cal E}}
    \newcommand{\FF}{{\cal F}}
    \newcommand{\HH}{{\cal H}}
    \newcommand{\II}{{\cal I}}
    \newcommand{\KK}{{\cal K}}
    \newcommand{\LL}{{\cal L}}
    \newcommand{\MM}{{\cal M}}
    \newcommand{\NN}{{\cal N}}
    \newcommand{\CC}{{\cal C}}
    \newcommand{\PP}{{\cal P}}
    \newcommand{\QQ}{{\cal Q}}
    \newcommand{\RR}{{\cal R}}
    \newcommand{\UU}{{\cal U}}
    \newcommand{\YY}{{\cal Y}}
    \newcommand{\SOSO}{{\cal SO}}
    \newcommand{\GLGL}{{\cal GL}}

  \newcommand{\NNN}{{\mathbb{N}}}
  \newcommand{\ZZZ}{{\mathbb{Z}}}
  \newcommand{\RRR}{{\mathbb{R}}}
  \newcommand{\CCC}{{\mathbb{C}}}
  \newcommand{\one}{{{\bf 1}}}

  \newcommand{\<}{{\ensuremath\langle}}
  \renewcommand{\>}{{\ensuremath\rangle}}
  \newcommand{\Aut}{{\rm Aut}}
  \newcommand{\cor}{{\rm cor}}
  \newcommand{\corr}{{\rm corr}}
  \newcommand{\cov}{{\rm cov}}
  \newcommand{\sd}{{\rm sd}}
  \newcommand{\tr}{{\rm tr}}
  \newcommand{\lag}{\mathcal{L}}
  \newcommand{\inn}{\rfloor}
  \newcommand{\lie}{\pounds}
  \newcommand{\longto}{\longrightarrow}
  \newcommand{\speer}{\parbox{0.4ex}{\raisebox{0.8ex}{$\nearrow$}}}
  \renewcommand{\dag}{ {}^\dagger }
  \newcommand{\h}{{}^\star}
  \newcommand{\w}{\wedge}
  \newcommand{\too}{\longrightarrow}
  \newcommand{\To}{\Rightarrow}
  \newcommand{\Too}{\;\Longrightarrow\;}
  \newcommand{\ow}{\stackrel{\circ}\wedge}
  \newcommand{\feed}{\nonumber \\}
  \newcommand{\comma}{\; , \quad}
  \newcommand{\period}{\; . \quad}
  \newcommand{\del}{\partial}
  \newcommand{\point}{$\bullet~~$}
  \newcommand{\doubletilde}{
  ~ \raisebox{0.3ex}{$\widetilde {}$} \raisebox{0.6ex}{$\widetilde {}$} \!\!
  }
  \newcommand{\topcirc}{\parbox{0ex}{~\raisebox{2.5ex}{${}^\circ$}}}
  \newcommand{\sym}{\topcirc}

  \newcommand{\half}{\frac{1}{2}}
  \newcommand{\third}{\frac{1}{3}}
  \newcommand{\fourth}{\frac{1}{4}}

  \renewcommand{\_}{\underset}
  \renewcommand{\^}{\overset}

  \renewcommand{\small}{\footnotesize}
}

\newcommand{\argmax}[1]{\text{arg}\underset{#1}\max}
\newcommand{\argmin}[1]{\text{arg}\underset{#1}\min}
\newcommand{\kld}[2]{D\!\left(\,#1\,|\!|\,#2\,\right)}

\newcommand{\pathmt}{./}
\newcommand{\basepath}{./}
\newcommand{\setpath}[1]{\renewcommand{\pathmt}{#1}\renewcommand{\basepath}{#1}}
\newcommand{\pathinput}[2]{
  \renewcommand{\pathmt}{\basepath #1}
  \input{\pathmt #2} \renewcommand{\pathmt}{\basepath}}

\newcommand{\hide}[1]{[\small #1 \normalsize]}
\newcommand{\color}[2][1]{}

\article{10}{1}

\newcommand{\dimskip}{}

\title{\Large\textbf{A neural model for multi-expert architectures}}

\author{\normalsize Marc Toussaint\\
 \sizeix Institut f\"ur Neuroinformatik, Ruhr-Universit\"at Bochum\\
\sizeix 44780 Bochum, Germany\\
\textit{\sizeix Marc.Toussaint@neuroinformatik.ruhr-uni-bochum.de}}

\date{}

\begin{document}


\twocolumn[\mytitle]\thispagestyle{fancy}

\rhead{\it Proceedings of the International Joint Conference on Neural Networks (IJCNN 2002)}

\begin{abstract}%
  We present a generalization of conventional artificial neural
  networks that allows for a functional equivalence to multi-expert
  systems. The new model provides an architectural freedom going
  beyond existing multi-expert models and an integrative formalism to
  compare and combine various techniques of learning. (We consider
  gradient, EM, reinforcement, and unsupervised learning.) Its uniform
  representation aims at a simple genetic encoding and evolutionary
  structure optimization of multi-expert systems. This paper contains
  a detailed description of the model and learning rules, empirically
  validates its functionality, and discusses future perspectives.
\end{abstract}

\section{Introduction}

When using multi-expert architectures for modeling behavior or data,
the motivation is the separation of the stimulus or data space into
disjoint regimes one which separate models (experts) are applied
\cite{jacobs:99,jacobs:90}. The idea is that experts responsible for
only a limited regime can be smaller and more efficient, and that
knowledge from one regime should not be extrapolated onto another
regime, i.e., optimization on one regime should not interfere with
optimization on another. Several arguments indicate that this kind of
adaptability cannot be realized by a single conventional neural
network \cite{toussaint:02}. Roughly speaking, for conventional neural
networks the optimization of a response in one regime always
interferes with responses in other regimes because they depend on the
same parameters (weights), which are not separated into disjoint
experts.

To realize a seperation of the stimulus space one could rely on the
conventional way of implementing multi-experts, i.e., allow neural
networks for the implementation of expert modules and use external,
often more abstract types of gating networks to organize the
interaction between these modules. Much research is done in this
direction \cite{bengio:94,cacciatore:94,jordan:94,rahman:99,ronco:97}.
The alternative we want to propose here is to introduce a neural model
that is capable to represent systems that are functionally equivalent
to multi-expert systems within a single integrative network. This
network does not explicitly distinguish between expert and gating
modules and generalizes conventional neural networks by introducing a
counterpart for gating interactions. What is our motivation for such a
new representation of multi-expert systems?
\begin{itemize}
\item First, our representation allows much more and qualitatively new
  architectural freedom. E.g., gating neurons may interact with expert
  neurons; gating neurons can be a part of experts. There is no
  restriction with respect to serial, parallel, or hierarchical
  architectures---in a much more general sense as proposed in
  \cite{jordan:94}.

\item Second, our representation allows in an intuitive way to combine
  techniques from various learning theories. This includes gradient
  descent, unsupervised learning methods like Hebb learning or the Oja
  rule, and an EM-algorithm that can be transferred from classical
  gating-learning theories \cite{jordan:94}. Further, the
  interpretation of a specific gating as an action exploits the realm
  of reinforcement learning, in particular Q-learning and (though not
  discussed here) its TD and TD($\l$) variants \cite{sutton:98}.

\item Third, our representation makes a simple genetic encoding of
  such architectures possible. There already exist various techniques
  for evolutionary structure optimization of networks (see
  \cite{yao:99} for a review). Applied on our representation, they
  become techniques for the evolution of multi-expert architectures.

\end{itemize}

After the rather straight-forward generalization of neural
interactions necessary to realize gatings (section \ref{def}), we will
discuss in detail different learning methods in section \ref{learn}.
The empirical study in section \ref{emp} compares the different
interactions and learning mechanisms on a test problem similar to the
one discussed by Jacobs et al. \citeyear{jacobs:90}.

\section{Model definition}\label{def}

\paragraph{Conventional multi-expert systems.}
Assume the system has to realize a mapping from an input space $X$ to
an output space $Y$. Typically, an $m$-expert architecture consists of
a gating function $\hat g:\; X \to [0,1]^m$ and $m$ expert functions
$f_i:\, X \to Y$ which are combined by the softmax linear combination:
\dimskip
\begin{align}
y&= \sum_{i=1}^m g_i\, f_i(x) \;,\quad
g_i=\frac{e^{\b \hat g_i(x)}}{\sum_{j=1}^m e^{\b \hat g_j(x)}} \;,
\dimskip
\end{align}
where $x$ and $y$ are input and output, and $\b$
describes the ``softness'' of this winner-takes-all type competition
between the experts, see Figure \ref{multiexp}. The crucial question
becomes how to train the gating. We will discuss different methods in
the next section.

\paragraph{Neural implementation of multi-experts.}
We present a single neural system that has at least the capabilities
of a multi-expert architecture of several neural networks. Basically
we provide additional competitive and gating interactions, for an
illustration compare Figure \ref{multiexp} and Figure \ref{samples}-B.
More formally, we introduce the model as follows:

The architecture is given by a directed, labeled graph of neurons
$(i)$ and links $(ij)$ from $(j)$ to $(i)$, where $i,j=1..n$. Labels
of links declare if they are ordinary, competitive or gating
connections. Labels of neurons declare their type of activation
function. With every neuron $(i)$, an activation state (output value)
$z_i\in[0,1]$ is associated. A neuron $(i)$ collects two terms of
excitation $x_i$ and $g_i$ given by \dimskip
\begin{align}
& x_i=\sum_{(ij)} w_{ij} z_j + w_i \label{xexi}\\
& g_i=
\left\{
{1 \atop \frac{1}{N_i}\, \sum_{(ij)^g} z_j}\quad
{\text{if }N_i = 0 \atop \text{else}}
\right.\comma N_i=\sum_{(ij)^g} 1 \;, \label{gexi}
\dimskip
\end{align}
where $w_{ij},w_i \in \RRR$ are weights and bias associated with the
links $(ij)$ and the neuron $(i)$, respectively. The second excitatory
term $g_i$ has the meaning of a gating term and is induced by
$N_i$ $g$-labeled links $(ij)^g$.

In case there are no $c$-labeled links $(ij)^c$ connected to a neuron
$(i)$, its state is given by
\dimskip
\begin{align}
  z_i = \p(x_i)\, g_i \;.
\label{stateSimple}
\dimskip
\end{align}
Here, $\p:\, \RRR \to [0,1]$ is a sigmoid function. This means, if a
neuron $(i)$ has no gating links $(ij)^g$ connected to it, then
$g_i=1$ and the sigmoid $\p(x_i)$ describes its activation.
Otherwise, the gating term $g_i$ multiplies to it.

Neurons $(i)$ that are connected by (bi-directed) $c$-labeled links
$(ij)^c$ form a \emph{competitive group} in which only one of the
neurons (the \emph{winner}) acquires state $z_{\text{winner}}=1$ while
the other's states are zero.  Let $\{i\}^c$ denote the competitive
group of neurons to which $(i)$ belongs. On such a group, we introduce
a normalized distribution $y_i$, $\sum_{j\in\{i\}^c} y_j=1$, given by
\dimskip
\begin{align}
  y_i = \frac{\psi(x_i)}{X_i} \;,\quad X_i=\sum_{k \in \{i\}^c}
  \psi(x_k) \;.
\label{normalization}
\dimskip
\end{align}
Here, $\psi$ is some function $\RRR \to \RRR$ (e.g., the exponential
$\psi(x)=e^{\b x}$). The neurons states $z_j\in\{0,1\}$, $j\in\{i\}^c$
depend on this distribution $y_i$ by one of the following competitive
rules of winner selection: We will consider a selection with
probability proportional to $y_i$ (softmax), deterministic selection
of the maximum $y_i$, and $\e$-greedy selection (where with
probability $\e$ a random winner is selection instead of the maximum).

\begin{figure}[t]\center
\includegraphics[bb=0 1 -2 0]{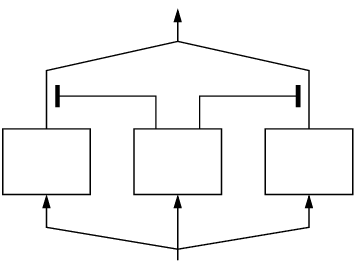}
\setlength{\unitlength}{2763sp}%
\begingroup\makeatletter\ifx\SetFigFont\undefined%
\gdef\SetFigFont#1#2#3#4#5{%
  \reset@font\fontsize{#1}{#2pt}%
  \fontfamily{#3}\fontseries{#4}\fontshape{#5}%
  \selectfont}%
\fi\endgroup%
\begin{picture}(2424,1749)(889,-973)
\put(1201,-413){\makebox(0,0)[b]{\smash{\SetFigFont{8}{9.6}{\rmdefault}{\mddefault}{\updefault}{\color[rgb]{0,0,0}$f_1$}%
}}}
\put(3001,-413){\makebox(0,0)[b]{\smash{\SetFigFont{8}{9.6}{\rmdefault}{\mddefault}{\updefault}{\color[rgb]{0,0,0}$f_2$}%
}}}
\put(3001,-226){\makebox(0,0)[b]{\smash{\SetFigFont{7}{8.4}{\rmdefault}{\mddefault}{\updefault}{\color[rgb]{0,0,0}expert}%
}}}
\put(1201,-226){\makebox(0,0)[b]{\smash{\SetFigFont{7}{8.4}{\rmdefault}{\mddefault}{\updefault}{\color[rgb]{0,0,0}expert}%
}}}
\put(2101,-226){\makebox(0,0)[b]{\smash{\SetFigFont{7}{8.4}{\rmdefault}{\mddefault}{\updefault}{\color[rgb]{0,0,0}gating}%
}}}
\put(2101,-410){\makebox(0,0)[b]{\smash{\SetFigFont{8}{9.6}{\rmdefault}{\mddefault}{\updefault}{\color[rgb]{0,0,0}$g$}%
}}}
\end{picture}
\setlength{\unitlength}{1pt}
\caption{Ordinary multi-expert architecture. Gating and experts modules are
  explicitly separated and the gating may not depend on internal
  states or the output of experts.}
\label{multiexp}
\end{figure}

\begin{figure*}[t]\center
\includegraphics[scale=0.5]{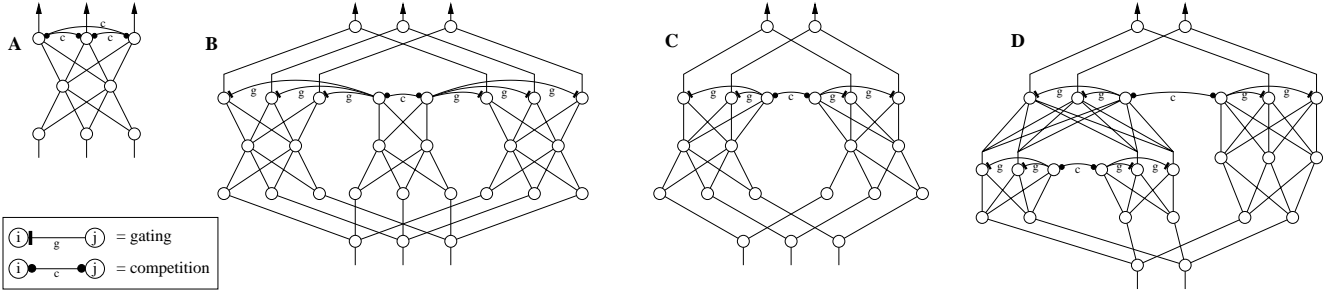}
\caption{Sample architectures. Please see section \ref{def} for a description of these
  architectures.}
\label{samples}
\end{figure*}

Please see Figure \ref{samples} to get an impression of the
architectural possibilities this representations provides. Example A
realizes an ordinary feed-forward neural network, where the three
output neurons form a competitive group. Thus, only one of the output
neurons will return a value of $1$, the others will return $0$.
Example $B$ realizes exactly the same multi-expert system as depicted
in Figure \ref{multiexp}. The two outputs of the central module form a
competitive group and gate the output neurons of the left and right
module respectively---the central module calculates the gating whereas
the left and right modules are the experts.  Example C is an
alternative way of designing multi-expert systems.  Each expert module
contains an additional output node which gates the rest of its outputs
and competes with the gating nodes of the other experts. Thus, each
expert estimates itself how good it can handle the current stimulus
(see the Q-learning method described below). Finally, example D is a
true hierarchical architecture. The two experts on the left compete to
give an output, which is further processed and, again, has to compete
with the larger expert to the right. In contrast, Jordan \& Jacobs
\citeyear{jordan:94} describe an architecture where the calculation of
one single gating (corresponding to only one competitive level) is
organized in a hierarchical manner. Here, several gatings on different
levels can be combined in any successive, hierarchical way.

\section{Learning}\label{learn}

In this section we introduce four different learning methods, each of
which is applicable independent of the specific architecture. We
generally assume that the goal is to approximate training data given
as pairs $(x,t)$ of stimulus and target output value.

\paragraph{The gradient method.}
To calculate the gradient, we assume that selection in competitive
groups performed with probability proportional to the distribution
$y_i$.  We calculate an approximate gradient of the conditional
probability $\PP(y|x)$ that this system represents by replacing the
actual state $z_i$ in Eq.~(\ref{xexi}) by its expectation value
$y_i$ for neurons in competitive groups (see also \citeNP{neal:90}).
For the simplicity of notation, we identify $z_i\equiv y_i$. Then, for
a neuron $(i)$ in a competitive group obeying
Eq.~(\ref{normalization}), we get the partial derivatives of the
neuron's output with respect to its excitations: \dimskip
\begin{align}
\frac{\del z_i}{\del x_j}
 &=\frac{\psi'(x_i)\, \d_{ij}}{X_i} -
   \frac{\psi(x_i)}{(X_i)^2}\, \Big[\psi'(x_j)\, \d_{j\in\{i\}^c}\Big] \feed
 &=\frac{\psi'(x_j)}{X_i}\, \big[ \d_{ij} - z_i\, \d_{j\in\{i\}^c}
 \big] \;,
\label{partial1}\\
\frac{\del z_i}{\del g_j} &= 0 \;,\label{partial2}
\dimskip
\end{align}
where $\d_{j\in\{i\}^c}=1$ iff $j$ is a member of $\{i\}^c$. Let
$E=E(z_1,..,z_n)$ be an error functional. We write the delta-rule for
back-propagation by using the notations $\check \d_i=\frac{dE}{dz_i}$
and $\big(\d_i=\frac{dE}{d x_i},~\d^g_i=\frac{dE}{d g_i}\big)$ for the
gradients at a neuron's output and excitations, respectively, and
$e_i=\frac{\del E}{\del z_i}$ for the local error of a single (output)
neuron. From Eqs.
(\ref{xexi},\ref{gexi},\ref{partial1},\ref{partial2}) we get
\dimskip
\begin{align}
\check \d_i 
  = \frac{dE}{dz_i}
 &= e_i + \sum_{j} \frac{dE}{dx_j}\, \frac{\del x_j}{\del z_i}
        + \sum_{j} \frac{dE}{dg_j}\, \frac{\del g_j}{\del z_i} \feed
 &= e_i + \sum_{(ji)} \d_j\, w_{ji} +\sum_{(ji)^g} \d^g_j\,\frac{1}{N_i}\;,\label{locerr}
\\
\d_i
  = \frac{dE}{dx_i}
 &= \sum_j \check \d_j\, \frac{\del z_j}{\del x_i}
  = \frac{\psi'(x_i)}{X_i}\, 
    \big[ \check \d_i -\!\!\!\sum_{j\in\{i\}^c}\!\!\! \check \d_j\, z_j \big] \;,
\label{compDeltaI}\\
\d^g_i
  = \frac{dE}{d g_i}
 &= \sum_j \check \d_j\, \frac{\del z_j}{\del g_i} = 0 \;.
\dimskip
\end{align}
(In Eq. (\ref{compDeltaI}) we used $X_i=X_j$ for $i\in\{j\}^c$ and
$i\in\{j\}^c \Leftrightarrow j\in\{i\}^c$.) For neurons that do not
join a competitive group we get from Eq. (\ref{stateSimple})
\dimskip
\begin{align}
&\frac{\del z_i}{\del x_j}
  = \p'(x_i)\, g_i\, \d_{ij} \comma
 \frac{\del z_i}{\del g_j}
  = \p(x_i)\, \d_{ij} \;, \\
&\d_i
  = \frac{dE}{dx_i}
  = \sum_j \check \d_j\, \frac{\del z_i}{\del x_j}
  = \p'(x_i)\, g_i\, \check\d_i \;,\\
&\d^g_i
  = \frac{dE}{d g_i}
  = \sum_j \check \d_j\, \frac{\del z_i}{\del g_j}
  = \p(x_i)\, \check\d_i \;,
\dimskip
\end{align}
where $\check \d_i$ is given in Eq. (\ref{locerr}). The final
gradients are
\dimskip
\begin{align}
\frac{dE}{d w_i}
 &= \d_i \comma \frac{dE}{d w_{ij}} = \d_i\, z_j \;.
\dimskip
\end{align}

The choice of the error functional is free. E.g., it can be chosen as
the square error $E=\sum_i (z_i-t_i)^2,~ e_i=2(z_i-t_i)$ or as the
log-likelihood $E=\ln \prod_i z_i^{t_i}\, (1-z_i)^{t_i},~
e_i=\frac{t_i}{z_i} - \frac{1-t_i}{1-z_i}$, where in the latter case
the target are states $t_i\in\{0,1\}$.

\paragraph{The basis for further learning rules.}
For the following learning methods we concentrate on the question:
\emph{What target values should we assume for the states of neurons in
  a competitive group?} In the case of gradient descent,
Eq.~(\ref{locerr}) gives the answer. It actually describes a linear
projection of the desired output variance down to all system states
$z_i$---including those in competitions. In fact, all the following
learning methods will adopt the above gradient descent rules except
for a redefinition of $\check \d_i$ (or alternatively $\d_i$) in the
case of neurons $(i)$ in competitive groups. This means that neurons
``below'' competitive groups are adapted by ordinary gradient descent
while the local error \emph{at} competitive neurons is given by other
rules than gradient descent.  Actually this is the usual way for
adapting systems where neural networks are used as internal modules
and trained by back-propagation (e.g., see \citeNP{anderson:94}).

\paragraph{An EM-algorithm.}
We briefly review the basic ideas of applying an EM-algorithm on the
problem of learning gatings in multi-experts \cite{jordan:94}. The
algorithm is based on an additional, very interesting assumption: Let
the outcome of a competition in a competitive group $\{c\}$ be
described by the states $z_i \in \{0,1\}$, $\sum_{i\in\{c\}} z_i=1$ of
the neurons that join this group. Now, we assume that there exists a
\emph{correct} outcome $h_i\in\{0,1\}$, $\sum_{i\in\{c\}} h_i=1$.
Formally, this means to assume that the complete training data are
triplets $(x,h_i,t)$ of stimuli, competition states, and output
values.\footnote{More precisely, the assumption is that there exists a
  teacher system of same architecture as our system. Our system adapts
  free parameters $w_{ij},~ w_i$ in order to approximate this teacher
  system. The teacher system produces training data and, since it has
  the same architecture as ours, also uses competitive groups to
  generate this data. The training data would be complete if it
  included the outcomes of these competitions.} However, the
competition training data is unobservable or \emph{hidden} and must be
inferred by statistical means. Bayes' rule gives an answer on how to
infer an expectation of the hidden training data $h_i$ and lays the
ground for an EM-algorithm. The consequence of this assumption is that
now the $y_i$ of competitive neurons are supposed to approximate this
expectation of the training data $h_i$ instead of being free. For
simplification, let us concentrate on a network containing a single
competitive group; the generalization is straightforward.
\begin{itemize}
\item Our system represents the conditional probability of output
  states $z^o$ and competition states $z^c$, depending on the stimulus
  $x$ and parameters $\t=(w_{ij},w_i)$:
\dimskip
\begin{align}\label{proba}
\PP(z^o,z^c|x,\t) = \PP(z^c|x,\t)\, \PP(z^o|z^c,x,\t) \;.
\dimskip
\end{align}

\item (E-step) We use Bayes rule to infer the expected competition
  training data $h_i$ hidden in a training tuple $(x,\cdot,t)$, i.e.,
  the probability of $h_i$ when $x$ and $t$ are given.
\dimskip
\begin{align}
\PP(h_i|x,t) = \frac{\PP(t|h_i,x) \PP(h_i|x)}{\PP(t|x)}
  \dimskip
\end{align}
  Since these probabilities refer to the training (or teacher) system,
  we can only approximate them. We do this by our current
  approximation, i.e., our current system:
\dimskip
\begin{align}\label{infer}
\PP(h_i|x,t,\t)
&= \frac{\PP(t|h_i,x,\t)\, \PP(h_i|x,\t)}{\PP(t|x,\t)} \feed
&= \frac{\PP(t|h_i,x,\t)\, \PP(h_i|x,\t)}{\sum_{z^c} \PP(z^c|x,\t)\, \PP(t|z^c,x,\t)} \;.
  \dimskip
\end{align}

\item (M-step) We can now adapt our system. In the classical
  EM-algorithm, this amounts to maximizing the expectation of the
  log-likelihood (cp. Eq.~(\ref{proba}))
\dimskip
\begin{align}
E[l(\t')] = E[\ln \PP(h|x,\t') + \ln \PP(t|z^c,x,\t')] \;,
  \dimskip
\end{align}
  where the expectation is with respect to the distribution
  $\PP(h|x,t,\t)$ of $h$-values (i.e., depending on our inference of
  the hidden states $h$); and the maximization is with respect to
  parameters $\t$. This equation can be simplified further---but, very
  similar to the ``least-square'' algorithm developed by Jordan \&
  Jacobs \citeyear{jordan:94}, we are satisfied to have inferred an
  explicit \emph{desired} probability $\hat y_i=\PP(h_i=1|x,t,\t)$ for
  the competition states $z_i$ that we use to define a mean-square
  error and perform an ordinary gradient descent.
\end{itemize}

Based on this background we define the learning rule as follows and
with some subtle differences to the one presented in
\cite{jordan:94}.  Equation (\ref{infer}) defines the desired
probability $\hat y_i$ of the states $z_i$. Since we assume a
selection rule proportional to the distribution $y_i$, the values
$\hat y_i$ are actually target values for the distribution $y_i$. The
first modification we propose is to replace all likelihood measures
involved in Eq.~(\ref{infer}) by general error measures $E$: Let us
define \dimskip
\begin{align}\label{actionvalue}
Q_i(x):=1-E(x) \quad \text{if $(i)$ wins.}
\dimskip
\end{align}
Then, in the case of the likelihood error $E(x)=1-\PP(t|x,\t)$, we
retrieve $Q_i(x)=\PP(t|h_i=1,x,\t)$. Further, let
\dimskip
\begin{align}\label{statevalue}
V(x):=\sum_i Q_i(x)\, y_i(x).
\dimskip
\end{align}
By these definitions we may rewrite Eq.~(\ref{infer}) as
\dimskip
\begin{align}\label{infer2}
\hat y_i(x)
  = \frac{Q_i(x)\, y_i(x)}{V(x)}
  = \frac{Q_i(x)\, y_i(x)}{\sum_j Q_j(x)\, y_j(x)} \;.
\dimskip
\end{align}
However, this equation needs some discussion with respect to its
explicit calculation in our context---leading to the second
modification. Calculating $Q_j(x)$ for every $j$ amounts to evaluating
the system for every possible competition outcome. One major
difference to the algorithm presented in \cite{jordan:94} is that we
do not allow for such a separated evaluation of all experts in a
single time step. In fact, this would be very expensive in case of
hierarchically interacting competitions and experts because the
network had to be evaluated for each possible combinatorial state of
competition outcomes. Thus we propose to use an approximation: We
replace $Q_j(x)$ by its average over the recent history of cases where
$(j)$ won the competition, \dimskip
\begin{align}
\bar Q_j \gets \g\, \bar Q_j + (1-\g)\, Q_j(x) \quad\text{whenever $(j)$ wins}\;,
\dimskip
\end{align}
where $\g \in [0,1]$ is a trace constant (as a simplification of the
time dependent notation, we use the algorithmic notation $\gets$ for
a replacement if and only if $(j)$ wins). Hence, our adaptation rule
finally reads
\dimskip
\begin{align}\label{EMrule}
\check \d_i = -\a_c\, \bigg[
  y_i - \frac{Q_i\, y_i}{\sum_{j \in \{i\}^c} \bar Q_j\, y_j}\bigg]
  \qquad\text{if $(i)$ wins},
\dimskip
\end{align}
and $\check \d_i = 0$ if $(i)$ does not win; which means a gradient
descent on the squared error between the approximated desired
probabilities $\hat y_i$ and the distribution $y_i$.

\paragraph{Q-learning.}
Probably, the reader has noticed that we chose the notations in the
previous section in the style of reinforcement learning: If one
interprets the winning of neuron $(i)$ as a decision on an action,
then $Q_i(x)$ (called \emph{action-value function}) describes the
(estimated) quality of taking this decision for stimulus $x$; whereas
$V(x)$ (called \emph{state-value function}) describes the estimated
quality for stimulus $x$ without having decided yet, see
\cite{sutton:98}. In this context, Eq.~(\ref{infer2}) is very
interesting: it proposes to adapt the probability $y_i(x)$ according
to the ratio $Q_i(x)\big/V(x)$---the EM-algorithm acquires a very
intuitive interpretation. To realize this equation without the
approximation described above one has to provide an estimation of
$V(x)$, e.g., a neuron trained on this target value (a \emph{critic}).
We leave this for future research and instead directly address the
Q-learning paradigm.

For Q-learning, an explicit estimation of the action-values $Q_i(x)$
is modeled. In our case, we realize this by considering $Q_i(x)$ as
the target value of the excitations $x_i$, $i\in\{c\}$, i.e., we train
the excitations of competing neurons toward the action values,
\dimskip
\begin{align}\label{Qrule}
\d_i = \a_c\, \left\{{x_i - Q_i \atop 0} \quad {\text{if $(i)$ wins}
    \atop \text{else}}\right. \;.
\dimskip
\end{align}
This approach seems very promising---in particular, it opens the door
to temporal difference and TD($\l$) methods and
other fundamental concepts of reinforcement learning theory.

\begin{table}[t]\small
  \fbox{\begin{colpage}

    -- The adaptation rate is $\a=0.01$ for all algorithms (as indicated
    in Eqs.~(\ref{EMrule},\ref{Qrule},\ref{Orule}), the delta-values
    for neurons in competitive groups are multiplied by $\a_c$).

    -- Parameters are initialized normally distributed around zero
    with standard deviation $\s=0.01$.

    -- The sigmoidal and linear activation functions are
    $\p_s(x)=\frac{1}{1+\exp(-10x)}$ and $\p_l(x)=x$, respectively.

    -- The competition function $\psi$ for softmax competition is
    $\psi_s(x)=e^{5 x}$.

    -- The Q-learning algorithm uses $\e$-greedy selection with
    $\e=0.1$; the others select either the maximal activation or with
    probability proportional to the activation.

    -- The values of the average traces $\bar Q_i$ and $\bar V$ are
    initialized to $1$.

    -- The following parameters were used for the different learning
    schemes: {\center
\begin{tabular}{cccccc}
\hline
\hline
       & gradient &  EM  &  Q   & Oja-Q \\
\hline
$\a_c$ & -- & 1 & 10  & 100 \\
$\g$   & -- & 0.9 & -- & 0.9 \\
$\psi$ & $\psi_s$ & $\p_s$ & $\p_l$ & $\p_l$ \\
\small selection & \small proportional & max & greedy & max \\
\hline
\hline
\end{tabular}\\}
\medskip
Here, $\a_c$ is the learning rate factor, $\g$ is the average trace
parameter, and $\psi$ is the competition function.
\end{colpage}}
\caption{Implementation details\vspace{-4ex}}
\label{details}
\end{table}

\paragraph{Oja-Q learning.}
Besides statistical and reinforcement learning theories, also the
branch of unsupervised learning theories gives some inspiration for
our problem. The idea of hierarchically, serially coupled competitive
groups raises a conceptual problem: Can competitions in areas close to
the input be trained without functioning higher level areas (closer to
the output) and vice versa? Usually, back-propagation is the standard
technique to address this problem.  But this does not apply on either
the EM-learning or the reinforcement learning approaches because they
generate a direct feedback to competing neurons in any layer.
Unsupervised learning in lower areas seems to point a way out of this
dilemma. As a first approach we propose a mixture of unsupervised
learning in the fashion of the normalized Hebb rule and Q-learning.
The normalized Hebb rule (of which the Oja rule is a linearized
version) can be realized by setting $\d_i = -\a_c\, z_i$ for a neuron
$(i)$ in a competitive group (recall $z_i \in \{0,1\}$). The gradient
descent with respect to adjacent input links gives the ordinary $\D
w_{ij} \propto z_i\,z_j$ rule.  Thereafter, the input weights
(including the bias) of each neuron $(i)$, $i \in \{c\}$ are
normalized. We modify this rule in two respects. First, we introduce a
factor ($Q_i - \bar V)$ that accounts for the success of neuron $(i)$
being the winner. Here, $\bar V$ is an average trace of the feedback:
\dimskip
\begin{align}
\bar V \gets \g\, \bar V + (1-\g)\, Q_i(x) \quad\text{every time step}\;,
\dimskip
\end{align}
where $(i)$ is the winner. Second, in the case of failure, $Q_i <
\bar V$, we also adapt the non-winners in order to increase their
response on the stimulus next time. Thus, our rule reads
\dimskip
\begin{align}\label{Orule}
\d_i = -\a_c\,(Q_i-\bar V) \left\{ {z_i \atop z_i-0.5} \quad
{\text{if $Q_i \ge \bar V$} \atop \text{else}} \right.\;.
\dimskip
\end{align}

Similar modifications are often proposed in reinforcement learning
models \cite{barto:85,barto:87}. The rule investigated here is only a
first proposal; all rules presented in the excellent survey of
Diamantaras \& Kung \citeyear{diamantaras:96} can equally be applied and
are of equal interest but have not yet been implemented by the author.

\section{Empirical study}\label{emp}

\begin{figure}[t]\center
\includegraphics[bb=0 1 -3 0]{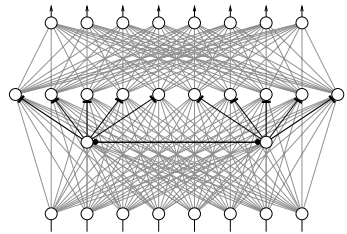}
\setlength{\unitlength}{1973sp}%
\begingroup\makeatletter\ifx\SetFigFont\undefined%
\gdef\SetFigFont#1#2#3#4#5{%
  \reset@font\fontsize{#1}{#2pt}%
  \fontfamily{#3}\fontseries{#4}\fontshape{#5}%
  \selectfont}%
\fi\endgroup%
\begin{picture}(4216,2874)(68,-1498)
\end{picture}
\setlength{\unitlength}{1pt}
\caption{The architecture we use for our experiments. All output
  neurons have linear activation functions $\p(x)=x$. All except the
  input neurons have bias terms.}
\label{testArch}
\end{figure}

We test the functionality of our model and the learning rules by
addressing a variant of the test presented in \cite{jacobs:90}. A
single bit of an 8-bit input decides on the subtask that the system
has to solve on the current input. The subtasks itself are rather
simple and in our case (unlike in \cite{jacobs:90}) are to map the
8-bit input either identically or inverted on the 8-bit output. The
task has to be learned online. We investigate the learning dynamics of
a conventional feed-forward neural network (FFNN) and of our model
with the 4 different learning methods. We use a fixed architecture
similar to an 8-10-8-layered network with 10 hidden neurons but
additionally install 2 competitive neurons that receive the input and
each gates half of the hidden neurons, see Figure \ref{testArch}. In
the case of the conventional FFNN we used the same architecture but
replaced all gating and competitive connections by conventional links.

\begin{figure}\center
\psfrag{error}{\small Error}
\includegraphics[scale=0.35]{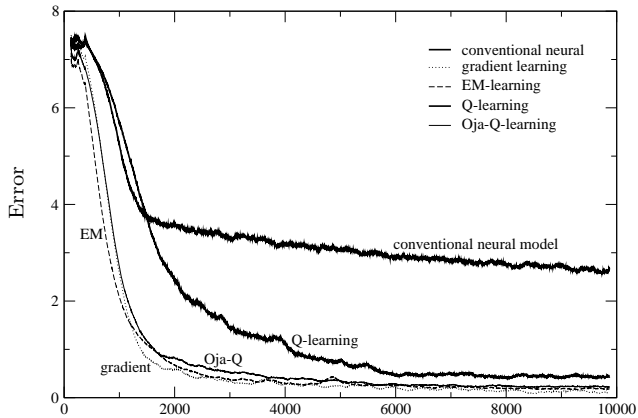}
\caption{Learning curves for the conventional neural network and the
  four different learning schemes.}
\label{curves}
\end{figure}

\begin{figure}\center
\input{psfrac}
\begin{tabular}{cc}
\small 1st subtask & \small 2nd subtask \\
\!\!\!\!\includegraphics[width=120pt,height=50pt]{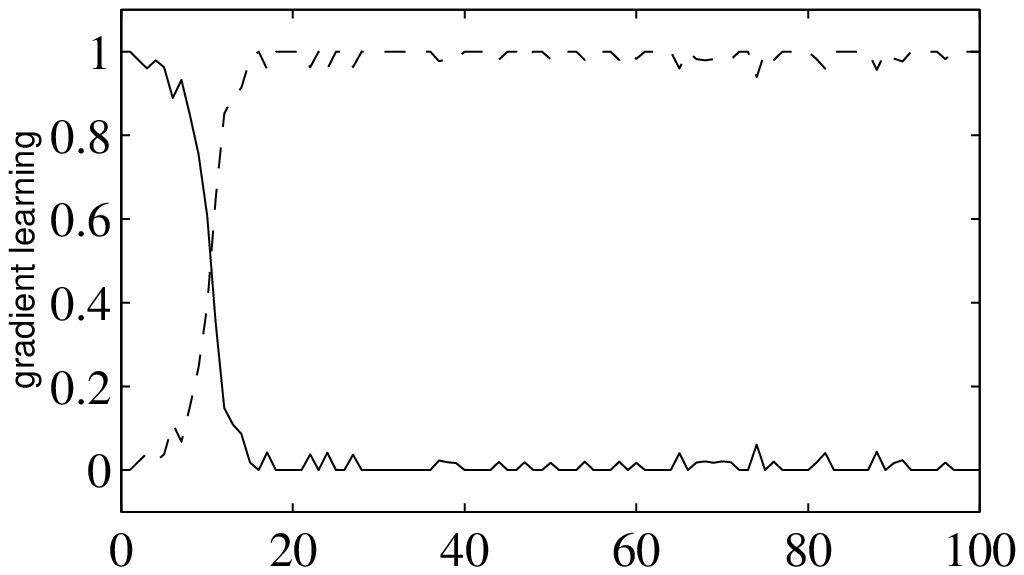}&
\!\!\!\!\includegraphics[width=120pt,height=50pt]{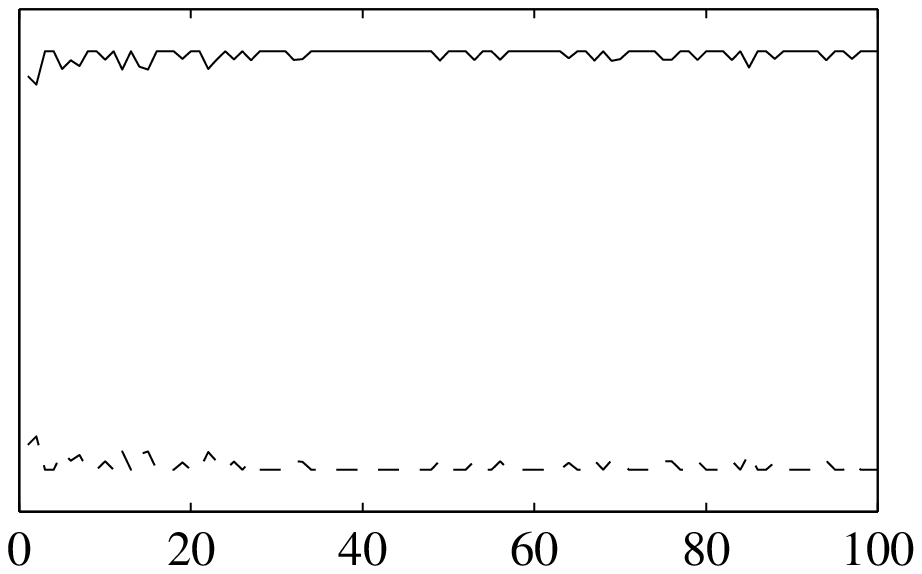}\\
\!\!\!\!\includegraphics[width=120pt,height=50pt]{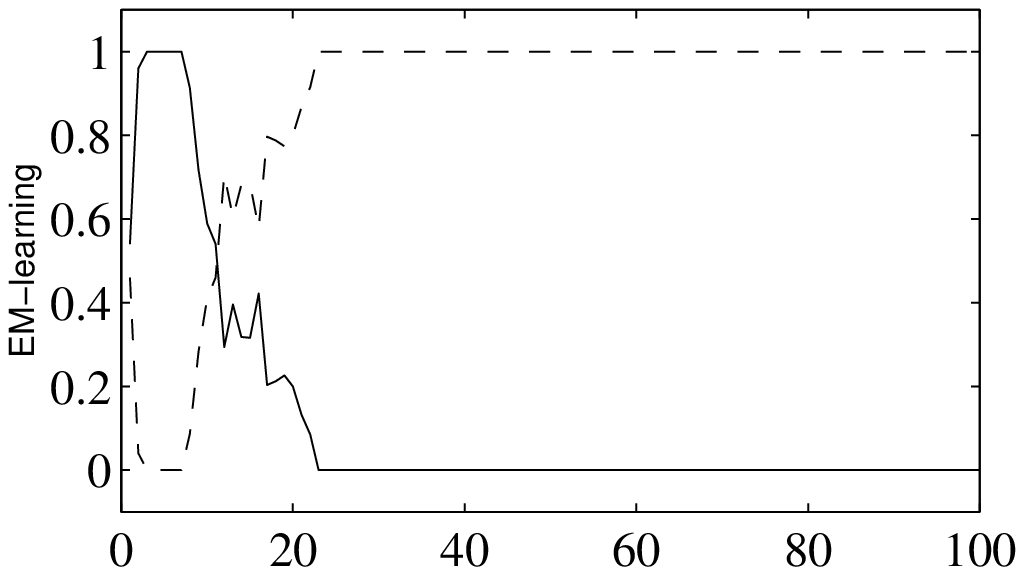}&
\!\!\!\!\includegraphics[width=120pt,height=50pt]{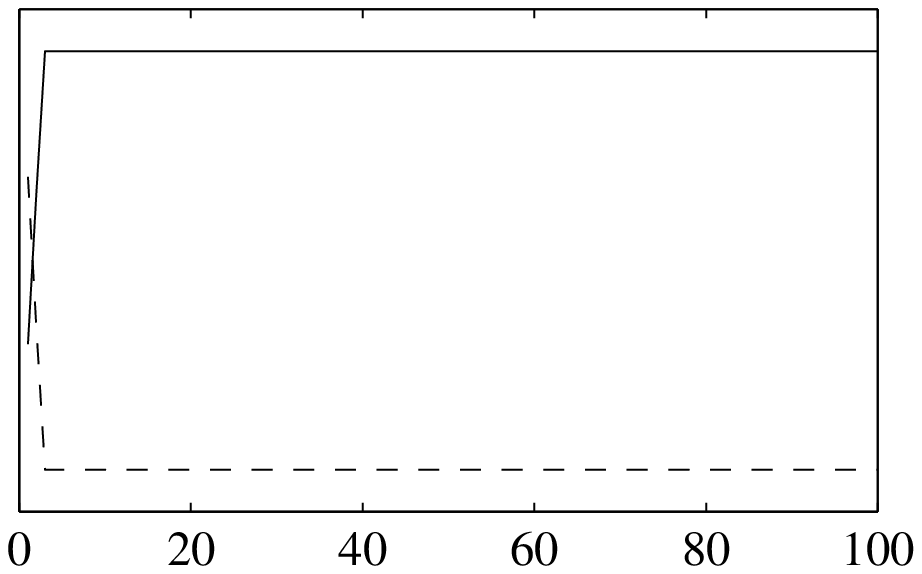}\\
\!\!\!\!\includegraphics[width=120pt,height=50pt]{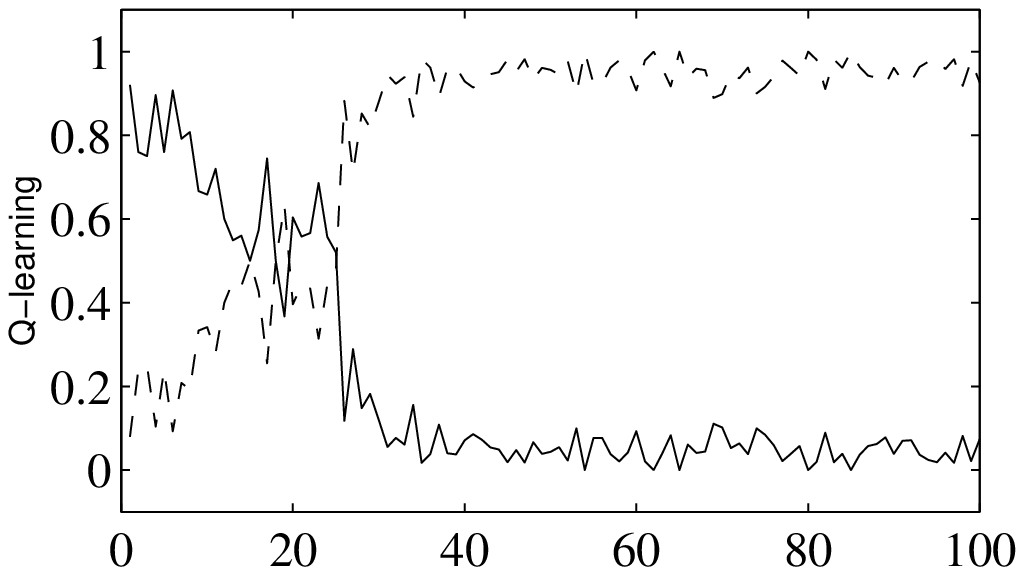}&
\!\!\!\!\includegraphics[width=120pt,height=50pt]{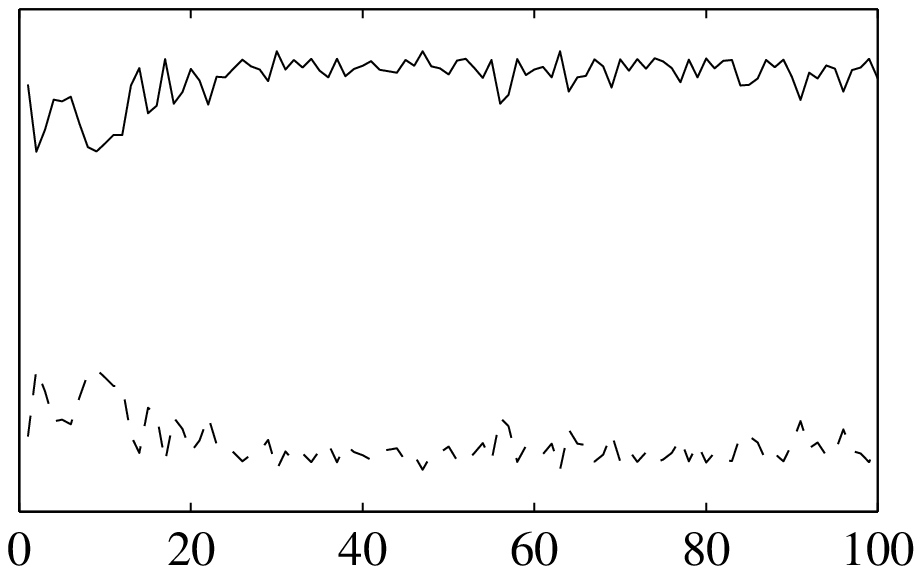}\\
\!\!\!\!\includegraphics[width=120pt,height=50pt]{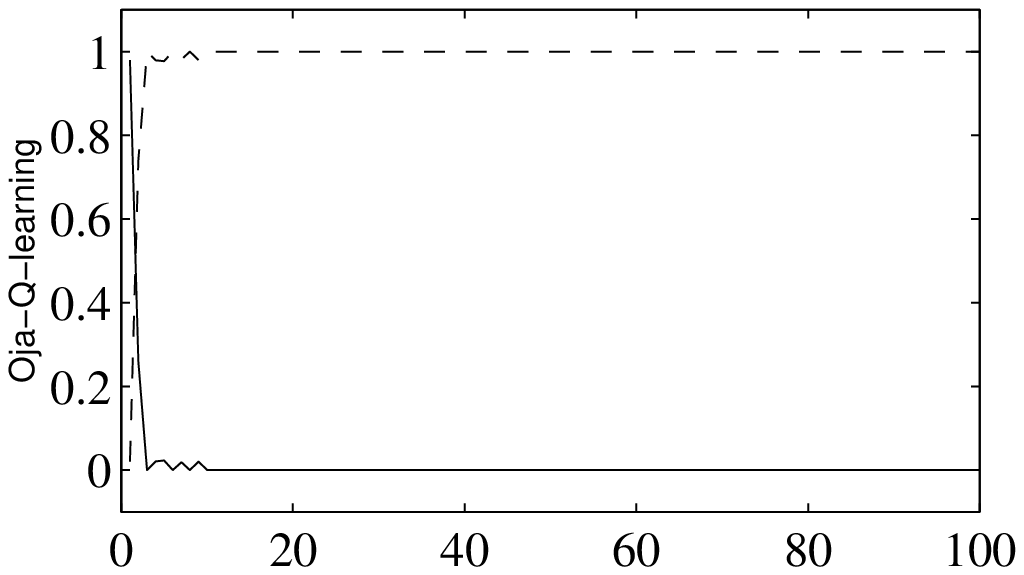}&
\!\!\!\!\includegraphics[width=120pt,height=50pt]{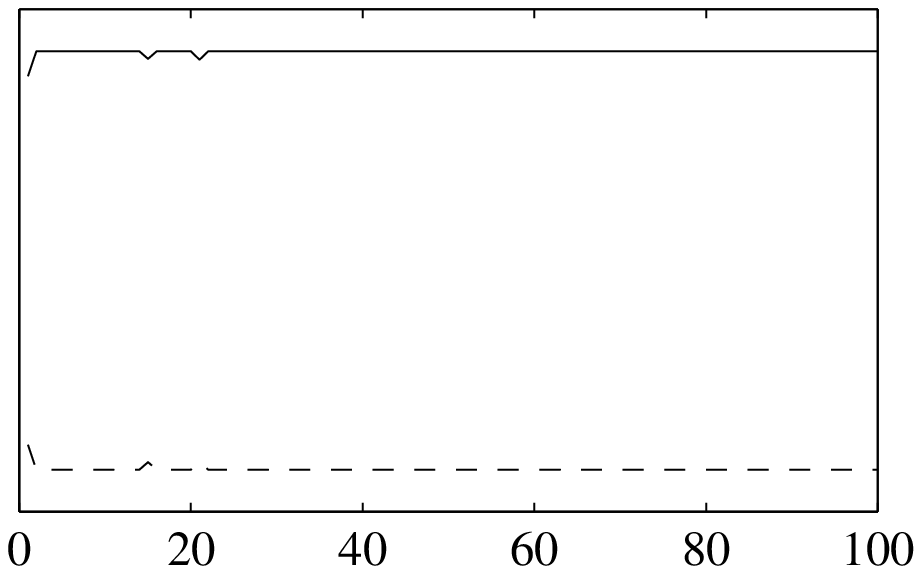}
\end{tabular}
\caption{The gating ratios for single trials for the four
  different learning schemes: The four rows refer to gradient, EM-,
  Q-, and Oja-Q-learning; and the two columns refer to the two classes
  of stimuli---one for the ``identical'' task, and one for the ``not''
  task. Each graph displays two curves that sum to 1 and indicate how
  often the first or second gating neuron wins in case of the
  respective subtask.}
\label{separation}
\end{figure}

Figure \ref{curves} displays the learning curves averaged over 20 runs
with different weight initializations. For implementation details see
Table \ref{details}. First of all, we find that all of the 4 learning
methods perform well on this task compared to the conventional FFNN.
The curves can best be interpreted by investigating if a task
separation has been learned. Figure \ref{separation} displays the
frequencies of winning of the two competitive neurons in case of the
different subtasks. The task separation would be perfect if these two
neurons would reliably distinguish the two subtasks. First noticeable
is that all 4 learning methods learn the task separation. In the case
of Q-learning the task separation is found rather late and remains
noisy because of the $\e$-greedy selection used. This explains its
slower learning curve in Figure \ref{curves}. EM and Oja-Q realize
strict task separations (maximum selection), for the gradient method
it is still a little noisy (softmax selection). It is clear that, if
the task separation has been found and fixed, all four learning
methods proceed equivalently. So it is no surprise that the learning
curves in Figure \ref{curves} are very similar except for a temporal
offset corresponding to the time until the task separation has been
found, and the non-zero asymptotic error corresponding to the noise of
task separation. (Note that Figure \ref{separation} represents only a
\emph{single}, typical trial.)

Generally, our experience was that the learning curves may look very
different depending on the weight initialization. It also happened
that the task separation was not found when weights and biases
(especially of the competing neurons) are initialized very large (by
$\NN(0,0.5)$). One of the competitive neurons then dominates from the
very beginning and prohibits the ``other expert'' to adapt in any way.
Definitely, a special, perhaps equal initialization of competitive
neurons could be profitable.

Finally, also the conventional FFNN only \emph{sometimes} solves the
task completely---more often when weights are initialized relatively
high. This explains the rather high error offset for its learning
curve.


\section{Conclusion}

We generalized conventional neural networks to allow for multi-expert
like interactions. We introduced 4 different learning methods for this
model and gave empirical support for their functionality. What makes
the model particularly interesting is:
\begin{enumerate}
\item The generality of our representation of system architecture
  allows new approaches for the structure optimization of multi-expert
  systems, including arbitrary serial, parallel, and hierarchical
  architectures. In particular evolutionary techniques of structure
  optimization become applicable.

\item The model allows the combination of various learning methods
  within a single framework. Especially the idea of integrating
  unsupervised learning methods in a system that adapts supervised
  opens new perspectives. Many more techniques from elaborated
  learning theories can be transfered on our model. In principle, the
  uniformity of architecture representation would allow to specify
  freely where it is learned by which principles.

\item The model overcomes the limitedness of conventional neural
  networks to perform task decomposition, i.e., to adapt in a
  decorrelated way to decorrelated data \cite{toussaint:02}.
\end{enumerate}

\subsection*{Acknowledgment}
I acknowledge support by the German Research Foundation DFG under
grant \emph{SoleSys}.

\small
\bibliography{/home/mt/bibtex/bibs}
\end{document}